\documentstyle[preprint,pra,aps,psfig,epsfig]{revtex}
\input{psfig.sty}

\begin{document}
\draft
\title{Dynamics of bright matter wave solitons in a quasi 1D
Bose-Einstein condensate with a
rapidly varying trap}
\author{F.Kh.\ Abdullaev$^{1,2}$ and R.\ Galimzyanov$^1$}
\address{$^1$Instituto de Fisica Teorica, Univrsidade Estadual Paulista
, Rua Pumplona, 145, Sao Paulo, Brasil}
\address{$^2$ Physical-Technical
Institute of Uzbek Academy of Sciences,700084, Tashkent-84, G. Mavlyanov
str., 2-b, Uzbekistan}

\date{\today}
\maketitle

\begin{abstract}
The dynamics of a bright matter wave soliton in a quasi 1D
Bose-Einstein condensate with periodically rapidly varying trap is
considered. The governing equation is derived based on averaging
over fast modulations of the Gross-Pitaevskii (GP) equation. This
equation has the form of GP equation with effective potential of
more complicated structure than unperturbed trap. For the case of
inverted (expulsive) quadratic trap corresponding to unstable  GP
equation, the effective potential can be stable. For the bounded
in space trap potential it is  showed that the bifurcation exists,
i.e.,the single well potential bifurcates to the triple well
effective potential.  Stabilization of BEC cloud on-site state
in the temporary modulated optical lattice is found. (analogous to
the Kapitza stabilization of the pendulum). The predictions of the
averaged GP equation are confirmed by the numerical simulations of
GP equation with rapid perturbations.

\end{abstract}

\newpage
\section{Introduction}
Recently bright matter wave solitons has been observed in a Bose-Einstein
condensate\cite{Hulet,Khaykovich}. In the
experiment described in Ref.\cite{Hulet} propagation in anisotropic BEC
with cigar type geometry was considered. In Ref. \cite{Khaykovich}
the soliton is monitored by projecting  the bound state of $\approx$5000 atoms
into expulsive harmonic potential. The soliton was observed propagating without
changing of the form for distances is order of $\sim 1$mm.
The expulsive harmonic trap used, in the experiment, corresponds to an unstable
potential for the GP
equation and a soliton can exists only as a metastable state
in BEC \cite{Carr,Nogami}. It is of interest to investigate the
dynamics of a bright matter wave soliton in inhomogeneous time
dependent systems, in particular possible stabilization of unstable
dynamics or complicated dynamics of solitons in BEC.

An interesting example of such a trap is an optical trap which is indeed
rapidly varied in time.
It is realized  for instance for
a condensate in dipole trap formed by a strong off-resonant laser
field \cite{Dum}. A typical model is a trap formed by a harmonic
potential with different frequencies which is cutting at an energy $V_c$.
Numerical simulations of the condensate dynamics and splits  in 1D BEC with a positive
scattering length under periodically shaken trap  has been performed in Ref.\cite{Dum}.
The same form of wave equation leads to the propagation of spatial
solitons in periodically modulated parabolic waveguides. This
problem appears in investigation of intense light beams in
a nonlinear waveguide with inhomogeneous distribution of
transverse refractive index. A typical distribution can be
approximated by a quadratic profile. The rapid variation of this
profile along the longitudinal direction leads to a rapidly varying quadratic
potential in the NLS equation.

The general mathematical problem is to investigate localized
states for the nonlinear wave equation with cubic nonlinearity and
with  rapidly varying (not small) potential of the form $V_0 =
f(x)\alpha(t/\epsilon), \epsilon \ll 1$. Also it is of interest to
investigate possibilities of stabilization of a soliton by rapid
perturbations. For multi dimensional NLS equation under rapidly
varying in {\it space} periodic potential such a problem is
discussed in \cite{Turytsin}.

The paper is organized as follows:

In section 2 we describe the procedure of obtaining the averaged
equation for the GP equation with an external potential, that is
rapidly varying in time and inhomogeneous in space, with the form
$ f(x)\cos(\Omega t)$,where $\Omega = 1/\epsilon$. In Section 3,
we considered some important potentials for applications, which
have the forms given by
 $f(x) = x^2 , 1/\cosh(bx)$, and $ \cos(kx) $.

\section{Averaged Gross-Pitaevskii equation}
The dynamics of BEC is described by the time dependent GP equation
\begin{equation}
i\psi_{\tau} = -\frac{\hbar^2}{2m}\Delta\psi + V(r,t)\psi + g_{0}|\psi|^2 \psi,
\end{equation}
where $m$ is the atom mass, $g = 4\pi\hbar^2 a_{s}/m$,with  $a_{s}$
the atomic scattering length. $a_{s}>0$ corresponds to the BEC with repulsive interaction
between atoms and $a_{s} < 0$ to the attractive interaction. The trap potential is given by
 $V =
m\omega^{2}(y^2 + z^2 )/2 + \alpha(t)(m\omega_{1}^{2}x^{2}/2 +
V_{1}(x,t))$, where $V_{1}(x,t)$ is bounded potential or the
optical lattice potential, $\alpha(t)$ describes the time
dependence of the potential. We will specify the form of $V_{1}$
later. Below we will consider the cigar type condensate with
$\omega^2 \gg \omega_{1}^2$. Within such restrictions we can look
for the solution of (1) with form $\psi(x,y,z,t) = R(y,z)U(x,t) $,
where $R$ satisfies  the equation $$-\frac{\hbar^2}{2m}\Delta R +
\frac{m\omega^2}{2}(y^2 + z^2 )R =\lambda_{\rho}R, \ \lambda =
\hbar\omega.$$

Averaging over the transverse mode $R$(i.e. multiplying by $R^{\ast}$, $$|R_{0}|^{2} = \frac{m\omega}{\pi\hbar}
\exp(-\frac{m\omega}{\hbar}\rho^2),$$
and
integrating over $\rho$) we obtain the quasi 1D GP for
$U$ \cite{Garcia}
\begin{equation}
i\hbar U_{\tau} = -\frac{\hbar^2}{2m}U_{xx} + (\frac{m\omega_{1}^{2}x^{2}}{2} + V_{1}(x,t))U + G|U|^{2}U,
\end{equation}
where $G =g\int|R|^{4}dxdy/\int|R|^{2}dxdy = (2\hbar|a_{s}|\omega) $.
In the dimensionless variables
$$t = \omega \tau/2,\  x = x/l,\  l =\sqrt{\frac{\hbar}{m\omega}},\
 u = \sqrt{2|a_{s}|}U,$$
we have the governing equation
\begin{equation}\label{1}
iu_t +  \beta u_{xx} + 2\sigma |u|^2 u = \alpha(t) f(x) u .
\end{equation}
where we will assume a periodic modulation of the trap
\begin{equation}
\alpha (t) = \alpha_{0} + \alpha_1 \sin(\Omega t).
\end{equation}
For example for $V = m\omega_{1}^{2}x^{2}/2, \alpha_{0}f(x) =
(\omega_{1}/\omega)^{2}x^{2}.$ Also  $\sigma = \pm 1$ are the
attractive  and repulsive two body interactions respectively. We
introduce the parameter $\beta =\pm 1$ to have possibility to use
the results for optical beam propagation. For BEC system $\beta =
1$.

The field $u(x,t)$ can be represented in the form of sum of slowly and
rapidly
varying parts $U(x,t)$ and $\xi(x,t)$
\begin{equation}\label{2}
u(x,t) = U(x,t) + \xi(x,t) .
\end{equation}

For obtaining the equation for an averaged field we will apply
the asymptotic procedure suggested in \cite{Turytsin}, namely we
will present the rapidly varying part of the field as  expansion on
 Fourier series
\begin{equation}\label{3}
\xi = A\sin(\Omega t) + B\cos(\Omega t) + C\sin(2\Omega t) + D\cos(2\Omega t)+ ...
\end{equation}
where $A, B, C, D$ are functions of (x,t) that are slowly varying in the scale $O(1)$
functions.
By substituting the equations (\ref{2}),(\ref{3}) into (\ref{1}) we
obtain the next set of equations for the slowly varying field and
the coefficients of the expansion for the rapidly varying component

\begin{eqnarray}\label{4}
iU_t + \beta U_{xx} + 2|U|^2 U + U^{*}(A^2 + B^2 + C^2 + D^2 ) + \nonumber\\
+ 2U(|A|^2 + |B|^2 + |C|^2 + |D|^2 ) + (BCA^{*} - |A|^2 D + ACB^{*} + |B|^2 D +
\nonumber \\
+ ABC^{*} + \frac{A^2 D^{*}}{2} + \frac{B^2 D^{*}}{2}) = \alpha f(x) U +
\frac{\alpha_1}{2} f(x) A \end{eqnarray}

\begin{eqnarray}\label{5}
iA_t - i\Omega B + \beta A_{xx}  + 4|U|^2 A + 2U^{*}(BC - AD) + 2U^2 A^{*} +
\nonumber \\
2U(-DA^{*} + CB^{*} + BC^{*} - AD^{*}) + (\frac{3}{2}|A|^2 A + ....) = \alpha_1 f(x) U -
\frac{\alpha_1}{2} f(x) D,
\end{eqnarray}

\begin{eqnarray}\label{6}
i\Omega A + iB_t + \beta B_{xx} + 4|U|^2 B + 2U^2 B^{*} + 2U^{*}(AC + BD) + ...
\nonumber \\
= \frac{\alpha_1}{2} f(x) C,
\end{eqnarray}

\begin{equation}\label{7}
iC_t - 2i\Omega D + \beta C_{xx} + 4|U|^2 C + 2U^{*} AB + ... =
\frac{\alpha_1}{2} f(x) B,
\end{equation}

\begin{equation}
2i\Omega C + iD_t + \beta D_{xx} + 4|U|^2 D + U^{*}(B^2 - A^2 ) +... =
-\frac{\alpha_1}{2}f(x) A.
\end{equation}

The parameters $\alpha_1 , \Omega$ are assumed $\gg 1$.
As we can find from this system, the coefficients can be
represented in the form of expansion
\begin{eqnarray}\label{8}
A = \frac{a_1}{\Omega^2} + \frac{a_2}{\Omega^4},\  B = \frac{b_1}{\Omega} +
 \frac{b_2}{\Omega^3}\nonumber \\
C = \frac{c_1}{\Omega^3} + \frac{c_2}{\Omega^5}, \ D = \frac{d_1}{\Omega^2} +
\frac{d_2}{\Omega^4}.
\end{eqnarray}
For the coefficients of expansion we have the expressions
\begin{eqnarray}\label{9}
a_1 = -i\alpha_1 f(x) U_t - \alpha_1 (f(x)U)_{xx}) - 2\alpha_1 f(x)
|U|^2 U + \alpha_0 \alpha_1 f^2 (x) U,\nonumber\\
b_1 = i\alpha_1 f(x) U, \ d_1 =
\frac{\alpha_1^2 f(x)^2}{4}U, c_1 =....
\end{eqnarray}

From (\ref{4}),(\ref{8}),(\ref{9}) we obtain the averaged equation
for $U$
\begin{eqnarray}\label{10}
iU_t + \beta U_{xx} + 2 |U|^2 U = \alpha_0 f(x)U -
i\frac{\epsilon^2}{2}f^2 (x)U_t
\nonumber \\
- \beta \frac{\epsilon^2}{2}f(x)(f(x)U)_{xx} -  2 \epsilon^2 f^2 (x)|U|^2 U +
 \frac{\epsilon^2 \alpha_0}{2}f^2 (x)U.
\end{eqnarray}

Here $\epsilon = \alpha_1 /\Omega$.

This equation has the conserved quantity - the number of atoms
\begin{equation}
\int_{-\infty}^{\infty}dx(1 + \frac{\epsilon^2 f^2 (x)}{2})|U|^2 = const.
\end{equation}

So, it is useful to introduce  the new field $V$ by
\begin{equation}\label{11}
V = (1 + \frac{\epsilon^2 f^2 (x)}{2})^{1/2}U.
\end{equation}

Substituting (\ref{11}) into (\ref{10}) and keeping the terms in order
$\epsilon^2 $
we obtain the equation
\begin{equation}\label{12}
iV_t + \beta V_{xx} + 2 |V|^2 V = (\alpha_0 f(x) +
\frac{\epsilon^2}{2} \beta [f_{x}(x)]^2 )V +
O(\epsilon^4).
\end{equation}
 The averaged equation has the form of a modified NLS equation with a
slowly varying potential
\begin{equation}\label{13}
W(x) = \alpha_{0} f(x) + \frac{\epsilon^2}{2} \beta [f_{x}(x)]^2 .
\end{equation}

This result shows that the soliton dynamics can have more
complicated character than in the case of BEC with slowly varying
parameters. We can expect here the stabilization of the unstable
dynamics of a soliton by rapidly varying perturbation. It is
a direct analogy with the stabilization of systems with a
few degrees of freedom under rapidly varying in time parameters.

\section{DYNAMICS OF DIFFERENT MODELS OF  TRAP POTENTIALS}
\subsection{A quadratic potential}
As an example, we will consider the attractive BEC with a
quadratic trap potential. In this case we have a perturbed trap
potential $[-\alpha_{0} + \alpha_{1}\sin(\Omega t)]f(x)$ with $f(x) =
x^{2}$ and $\beta = 1$ . An averaged effective potential is given by
\begin{equation}\label{quad}
W = (-\alpha_{0} + 2{\epsilon}^{2})x^2 ,
\end{equation}
where $\epsilon = \alpha_{1}/\Omega$. The case when
$\alpha_{0} > 0$ corresponds to the inverted (repulsive) harmonic
potential case which is well known to be unstable \cite{Nogami}.
As seen that when $2{\epsilon}^{2} > |\alpha_{0}| $ the sign of
the effective potential is reversed and the solitary wave dynamics
becomes stable. The frequency of oscillations of the soliton center now is
\begin{equation}\label{freq}
\Omega_{0} =2\sqrt{(2\epsilon^{2} - \alpha_{0})}.
\end{equation}

This result can be also obtained by using the moments method.
It is useful to introduce two new variables:
a position of a solitary solution of the governing equation (\ref{1})
\begin{equation}\label{X}
X(t) = \int_{-\infty}^{\infty}dx x|u(x,t)|^{2},
\end{equation}
and a field momentum
\begin{equation}\label{P}
P(t) = \int_{-\infty}^{\infty}dx (u^{*}(x,t)u_{x}(x,t) - u(x,t)u_{x}^{*}(x,t)).
\end{equation}
By using governing equation (\ref{1}) we come to the
following set of integro-differential equations :
\begin{eqnarray}\label{moment}
-iP_{t} + 2\int_{-\infty}^{\infty}dx'\frac{\partial W(x')}{\partial x'}
|u(x',t)|^2 = 0,\\
iX_{t} - \beta P = 0.\nonumber
\end{eqnarray}

In the case of a quadratic effective potential ($W(x) = \alpha x^2$) the set of
equations (\ref{moment}) is closed and we get the  equation of motion for the
soliton solution center, $X(t)$, in the form
\begin{equation}\label{moment1}
X_{tt} + 4\beta \alpha X = 0.
\end{equation}
We point out that the set (\ref{moment}) is exact. No additional
assumption was made in obtaining this set of equations, and it is
valid for any solitary solution of the governing equation
(\ref{1}). Also, it should be noted two remarkable facts connected
with the dynamics of a solitary solution of NLS equation in the
presence of a quadratic potential (or dynamics of BEC in a
quadratic trap): i).In this case the equations of motion
(\ref{moment}) for the soliton center, are exact, ii) after
averaging, the initially quadratic potential remains to be
quadratic but its sign may be reversed.

\begin{figure}
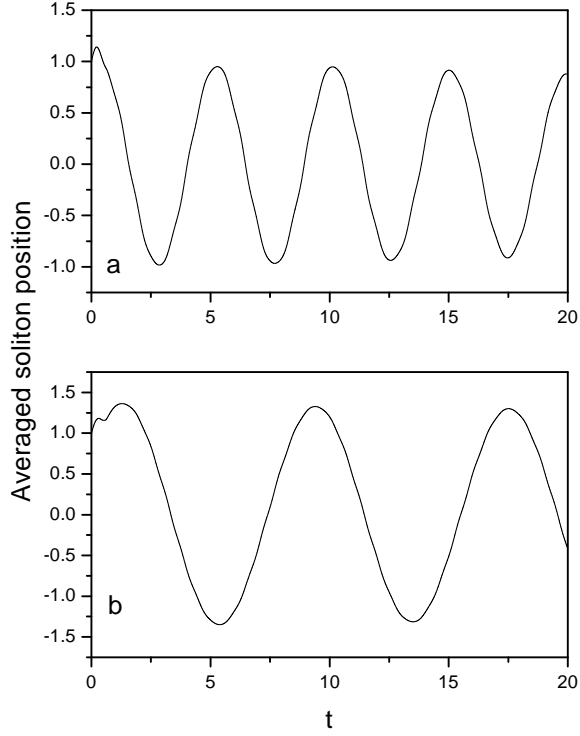

\caption{Numerical simulations of the full GP equation
for $f(x) = x^2$ with $\omega = 10$, $\alpha_{0} = 0.0493$ and
values of the perturbation parameter $\alpha_{1}$ providing reversing
of initially inverted harmonic potential.
Two cases are presented : a) $\alpha_{1} = 4.7124$ ; and
b) $\alpha_{1} = 3.1415$.
}
\label{fig_1}
\end{figure}

\begin{center}
\mbox{\hspace{-0.35in}
\psfig{figure=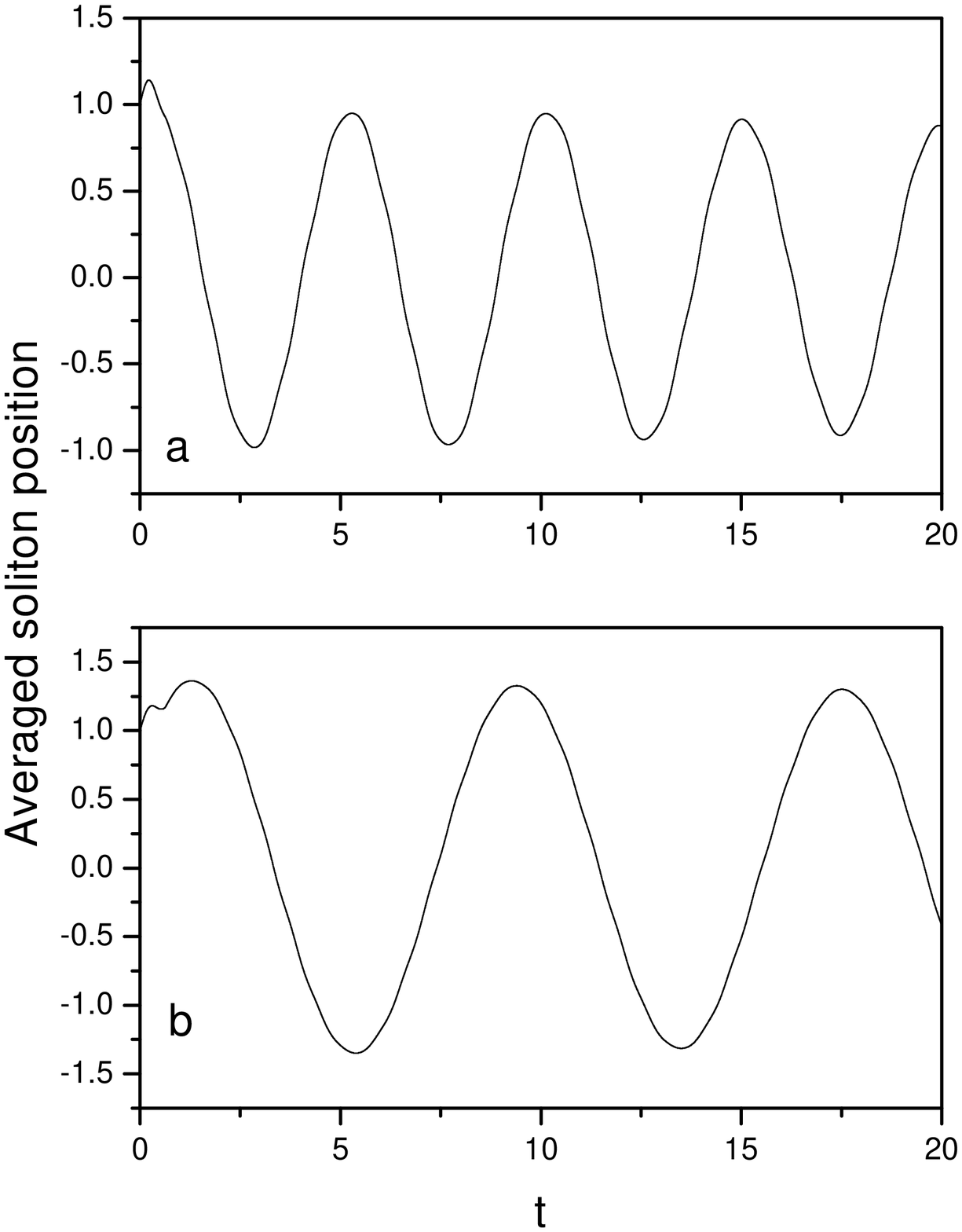,height=3.8in,angle=0.}}
\end{center}
  \vskip 0.2in
{\protect\small
Fig.  \ref{fig_1}. Numerical simulations of the full GP equation
for $f(x) = x^2$ with $\omega = 10$, $\alpha_{0} = 0.0493$ and
values of the perturbation parameter $\alpha_{1}$ providing reversing
of the initially inverted harmonic potential.
Two cases are presented : a) $\alpha_{1} = 4.7124$  and
b) $\alpha_{1} = 3.1415$.
}
\vskip0.2in

\noindent
The results of numerical simulation of the full Gross-Pitaevskii
equation with the modulated in time expulsive (inverted) harmonic potential
are shown in Fig. \ref{fig_1}. Two cases with $\alpha_{1} =
4.7124$ and $\alpha_{1} = 3.1415$ were considered. For both cases
$\alpha_{0} = 0.0493$ and  $\Omega = 10$. The results of
computation are the following : a) $\alpha_{1} = 4.7124$,  $\Omega
= 1.309, (T = 4.8)$ Calculation by Eq. (\ref{moment1}) gives
$\Omega_{cal} = 1.257, (T_{cal} = 5.0)$. b) $\alpha_{1} = 3.1415$,
$\Omega = 0.7705, (T = 8.155)$ Calculation by Eq. (\ref{moment1})
gives $\Omega_{cal} = 0.7695, (T_{cal} = 8.165)$.

In both cases we observe the appearance of  stable oscillations of
the soliton position. We obtain good  agreement between the full
simulations and the averaged equation. As seen, the greater is the
value of $\epsilon$,  the  discrepancy we found between results of
calculation of Eq.(\ref{moment1}) and the ones of full numerical
simulations of NLS equation. We can expect this since the ratio
$\alpha_{1}/\Omega$ is not small   for these values of parameters.
Let us estimate the values of parameters for the experimental
situation. In the experiment \cite{Khaykovich} the longitudinal
frequency was $\omega_{1} = 4$Hz, so, at the temporal modulations
of this potential, with the frequency $\Omega = 40$Hz we should
observe a stable bright matter wave soliton.

\subsection{The case of a bounded potential}
Let us consider a bounded in the space trap potential for the
atomic Bose-Einstein condensate \cite{Dum} under fast temporal
perturbations $(-\alpha_{0} + \alpha_{1}\sin(\Omega t))f(x)$ with
$f(x) = 1/\cosh(bx)$ and $\beta = 1$.

After averaging over the  period of fast perturbations,
 the effective potential $W(x)$ has the form
\begin{equation}\label{Veff}
W(x) = -\frac{\alpha_{0}}{\cosh(bx)} + \frac{{\epsilon}^{2}{b}^{2}}{2}
\frac{{\sinh(bx)}^2}{{\cosh(bx)^4}} .
\end{equation}
Provided that
\begin{equation}\label{cond}
\frac{{\epsilon}^{2}{b}^{2}}{2} > \frac{27}{2}\alpha_{0}
\end{equation}
this effective potential takes a triple well structure having a central
minimum and two lateral minima.
In Fig. \ref{fig_2}, a typical form of the potential is presented,
where $\alpha_{0} = 0.05, \alpha_{1} = 9, \omega = 10$ and $b = 0.6$.
(As seen, the condition (\ref{cond}) makes one to take large ratio of
parameters $\alpha_{1}$ and $\alpha_{0}$.)

\vskip 1.0in
\begin{center}
\mbox{\hspace{-0.35in}
\psfig{figure=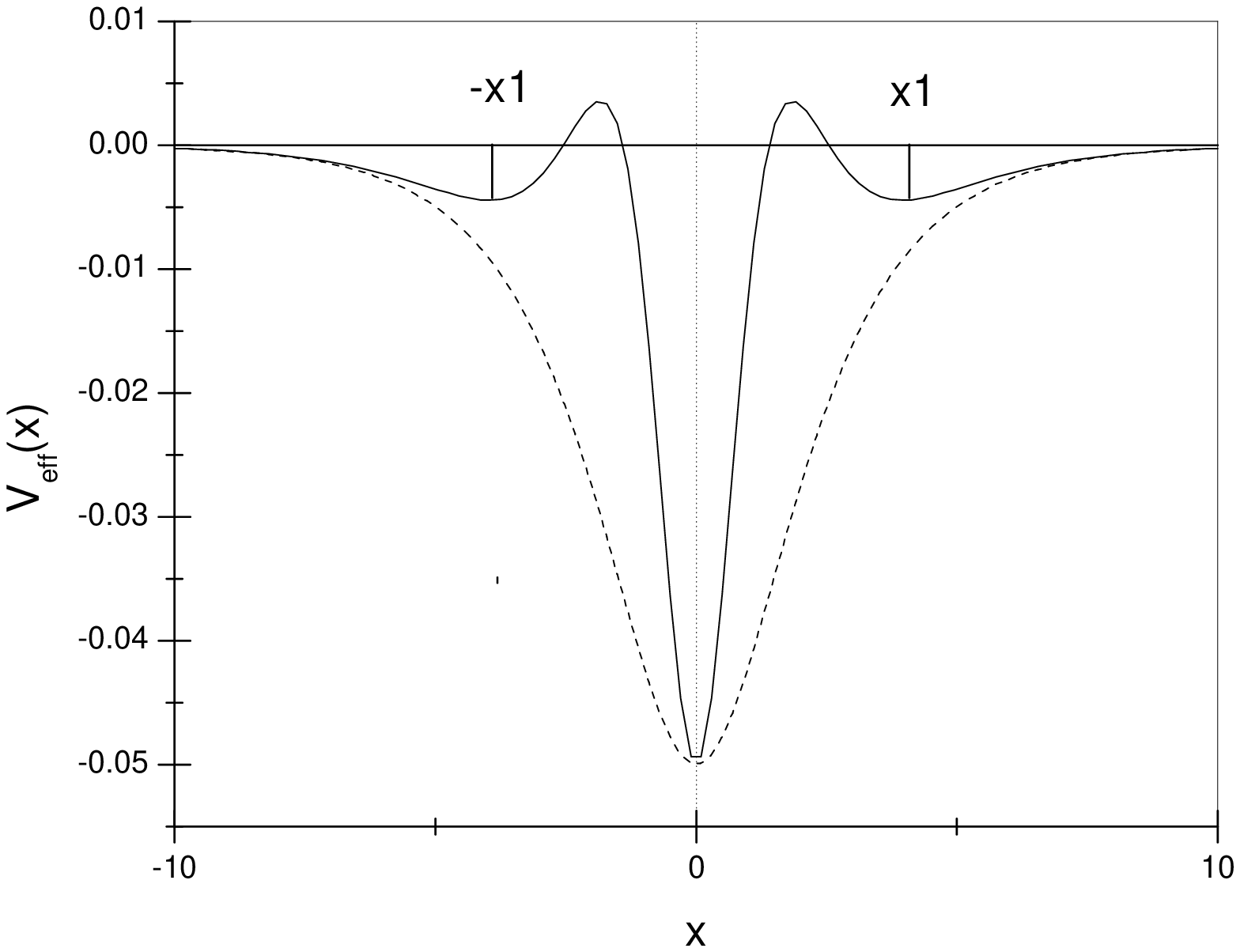,height=3.8in,angle=0.}}
\end{center}
  \vskip 0.2in
{\protect\small
Fig.  \ref{fig_2}.
Effective potential for $f(x) = 1/\cosh{(bx)}$ after
temporal averaging.
The parameters of perturbation $\omega = 10$, $\alpha_{0} = 0.05$ and
$\alpha_{1} = 9$.
Dashed line is for the case with $\alpha_{1} = 0$ when temporal
perturbation is turned off.}
\vskip0.2in

\noindent
The two lateral minima are positioned at  $x = \pm 3.96$ and the
central one at the point $x = 0$.

\begin{figure}
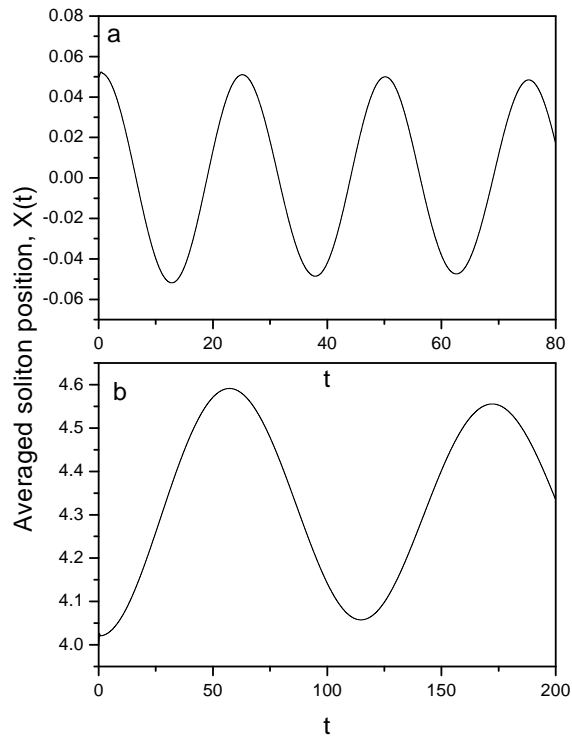

\caption{Numerical simulations of the full GP equation
for $f(x) = 1/\cosh(b x)$. The case a) corresponds to
the finite motion of the soliton center in the central minimum with
$X_{0} = 0$. The case b) corresponds to the motion in one of the lateral
minima with $X_{0} = 3.96$. For both cases
$\omega = 10$, $\alpha_{0} = 0.05$ and
$\alpha_{1} = 9$.
}
\label{fig_3}
\end{figure}

The results of numerical simulations  of the Gross-Pitaevskii equation are shown
in Fig. \ref{fig_3}.

\vskip 1.0in
\begin{center}
\mbox{\hspace{-0.35in}
\psfig{figure=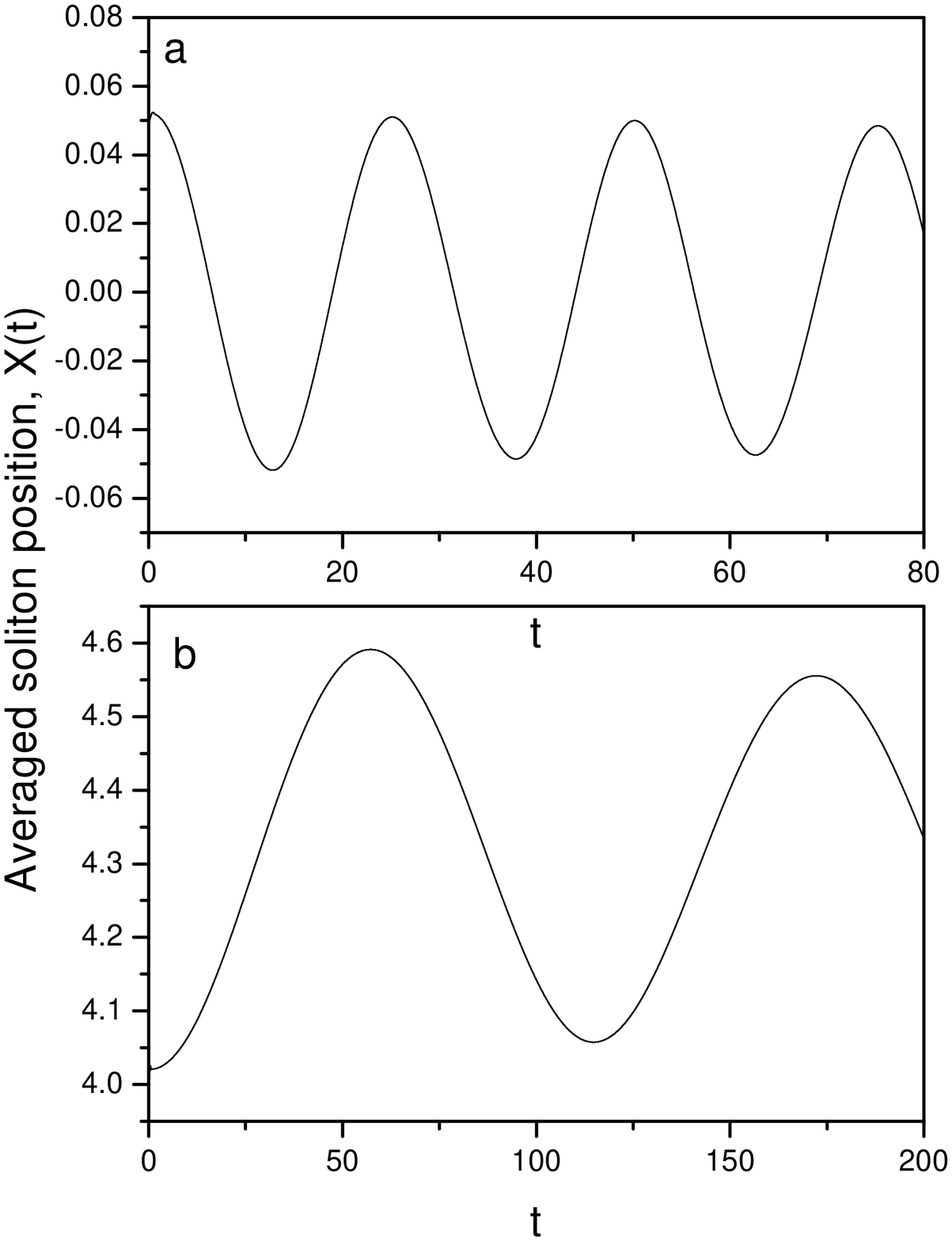,height=3.8in,angle=0.}}
\end{center}
  \vskip 0.2in
{\protect\small
Fig.  \ref{fig_3}.
Numerical simulations of the full GP equation
for the bounded potential with $f(x) = 1/\cosh(b x)$. The case a) corresponds to
the finite motion of the soliton center in the central minimum with
$X_{0} = 0$. The case b) corresponds to the motion in one of the lateral
minima with $X_{0} = 3.96$. For both cases
$\omega = 10$, $\alpha_{0} = 0.05$ and
$\alpha_{1} = 9$.
}
\vskip0.2in

\noindent

To describe small oscillations of the soliton solution center and
to obtain  equations of motion in general case, we will have to
make some approximations. As the ansatz, we suppose the solution
to take the form $u(x,t) = u(x-X(t),t)$. Here and after, for the
sake of simplicity, we also suppose the position of the minimum at
$X_{0} = 0$ and a symmetrical form for the soliton solution
(dynamics of which we are investigating)
\begin{equation}
|u(x,t)|^{2} = |u(-x,t)|^{2}.
\end{equation}
Now we can consider the second term of the equation (\ref{moment}).
At first we rewrite the effective potential as a sum of
antisymmetric and symmetric parts $W(x) = W^{A}(x) + W^{S}(x)$, where
$W^{A}(x) = \frac{1}{2}(W(x) - W(-x))$ and
$W^{S}(x) = \frac{1}{2}(W(x) + W(-x))$.
Then, substituting  $W(x)$ into the second term of Eq. (\ref{moment})
expanding $|u(x-X(t),t)|^{2}$ in terms of $X(t)$ and by holding the first two
terms,
\begin{equation}
|u(x'+X,t)|^{2} = |u(x,t)|^{2} + X\frac{\partial |u(x',t)|^{2}}{\partial x'} ,
\end{equation}
we have
\begin{eqnarray}\label{expan}
2\int_{-\infty}^{\infty}dx'\frac{\partial W(x')}{\partial x'}|u(x'+X,t)|^2 =
2\int_{-\infty}^{\infty}dx'\frac{\partial W^{A}(x')}{\partial x'}|u(x',t)|^2 +\\
2X\int_{-\infty}^{\infty}dx'\frac{\partial W^{S}(x')}{\partial x'}
\frac{\partial |u(x',t)|^2}{\partial x'} .
\end{eqnarray}
Introducing new parameters
$$\omega^2 = 2\beta \int_{-\infty}^{\infty}dx'
\frac{\partial W^{S}(x')}{\partial x'}
\frac{\partial |u(x',t)|^2}{\partial x'},\
\Delta X = \frac{\int_{-\infty}^{\infty}dx'\frac{\partial W^{A}(x')}
{\partial x'}|u(x',t)|^2}{\omega^2},$$
we get the following equation of motion for the center of soliton
\begin{equation}\label{moteq}
X_{tt} + \omega^2 (X + \Delta X) = 0.
\end{equation}
Calculation of the parameters $\Delta X$ and $\omega^2$ for our case
($\alpha_{0} = 0.05, \alpha_{1} = 9, \omega = 10$ and $b = 0.6$ with
$f(x) = 1/\cosh(bx)$) gives

1)  central minimum:  ${X_{0}}^W = 0$,   $\omega = 0.2539$ ($T = 24.74$),
$\Delta X = 0$, $X^{st} = 0$. Results of numerical simulation:
$\omega = 0.251$ (or $T = 25.0$), $\Delta X = 0$.

2)  a lateral minimum:  ${X_{0}}^W = 3.96$,   $\omega = 0.00599$ ($T = 104.73$),
$\Delta X = 0.321$, $X^{st} = {X_{0}}^W + \Delta X = 4.18$. Results of numerical simulation:
$\omega = 0.0539$ (or $T = 116.8$), $X^{st} = 4.3$.
Here ${X_{0}}^W$ is the position of a given minimum of the potential W(x);
${X_{0}}^{st}$ is the stationary point (about which oscillations occur);
$\Delta X$ is the shift of a stationary point;
$\omega$ and $T$ are the frequency and period of oscillations.

From the present analysis we conclude   that
the theory of averaging of GP equation over rapidly varying perturbations works
well in the case of bounded perturbation potential even when
the expansion parameter $\epsilon$ is sufficient large (in our case
$\epsilon = 0.9$). The effective trap potential bifurcates from the one well form to the
triple well structure. This can lead to the splitting of single attractive  BEC into
three parts.

\subsection{\bf A periodic potential}

Let $f(x) = \cos(kx)$. Then the averaged equation coincide with the unperturbed
one but with the another  potential:
\begin{equation}\label{e13}
F(x) = -\alpha_{0} \cos(kx) + \frac{\epsilon^2 k^2}{4}(1 - \cos(2kx)).
\end{equation}
 This potential has a more complicated structure and, as a result, the motion of
the soliton center has new properties.
Let us consider the motion of one soliton in such a potential.
The single soliton solution is
\begin{equation}\label{e14}
V(x,t) = 2i\eta \mbox{sech}[2\eta (x - \zeta )]\exp[i\frac{\xi z}{\eta} +
4i(\eta^{2} + \xi^{2})t +
i \delta],
\end{equation}
where $z = 2\eta(x-\zeta)$, and $\zeta = 4\xi t$ for the unperturbed
problem.
The effective potential acting on the soliton mass center is
\begin{equation}\label{e15}
W_{sol} = \frac{\pi \alpha k}{2\eta \sinh(\frac{\pi k}{2\eta} )}\cos(k\zeta) -
\frac{\pi \epsilon^2 k^3 }{2\eta \sinh(\frac{\pi k}{4\eta})}\cos(2k\zeta).
\end{equation}

The equation of motion  is obtained via
\begin{equation}
\zeta_{tt} = -\frac{\partial W_s}{\partial \zeta}.
\end{equation}
So, for the soliton center $\zeta$, we have the following
equation of motion
\begin{equation}\label{e16}
\zeta_{tt} = -\frac{\pi \epsilon^2 k^4}{2\eta \sinh(\frac{\pi
k}{2\eta})}\sin(2k\zeta ) + \frac{\pi k^2 \alpha}{2\eta
\sinh(\frac{\pi k}{4\eta})}\sin(k\zeta ).
\end{equation}
As seen the motion of the soliton center occurs under the action of {\it two
harmonic} effective perturbation. It represents a close  analogy with
the stabilization of a pendulum with the rapidly oscillating support
point (the Kapitza pendulum).  As it is
well known \cite{Landau}, the oscillations of the pendulum
at the  point $k\zeta = \pi$ are
unstable. The rapid perturbation stabilize this fixed point. The
stabilization condition follows from the effective  potential for
the soliton in the periodic field (\ref{e15}) and it is given by
\begin{equation}\epsilon^2 k^2 > \alpha
\cosh(\frac{k \pi}{4\eta}).
\end{equation}

\begin{figure}
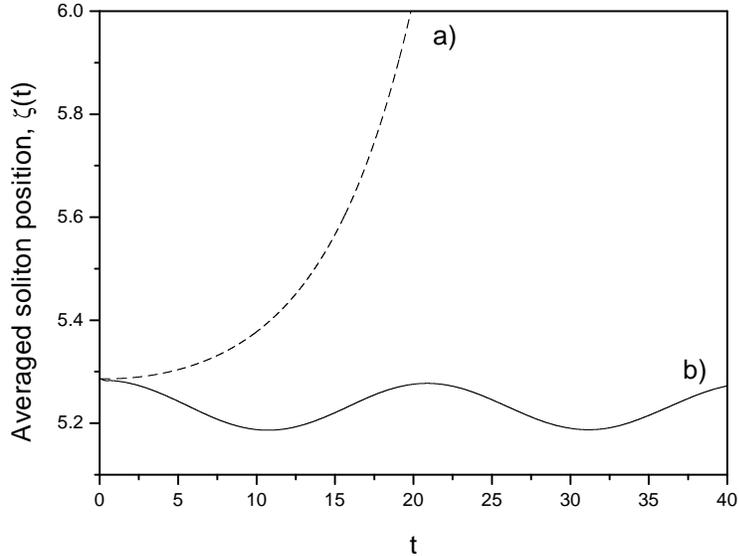

\caption{ Numerical simulations of the full GP equation for a
spatial periodic potential $f(x) = \cos(b x)$ with the initial
soliton position at an unstable point $X_{0} = \pi /k$. The case
a) (dashed line) corresponds to infinite motion of the soliton
center when $\alpha_{1} = 0$ and fast perturbation is turned off.
The case b) corresponds to $\alpha_{1} = 9$ when the fast
perturbation stabilizes the motion of the soliton center. The
parameters of perturbation $\omega = 10$, $\alpha_{0} = 0.05$ as
in the case of a bounded potential. } \label{fig_4}
\end{figure}

\begin{center}
\mbox{\hspace{-0.35in}
\psfig{figure=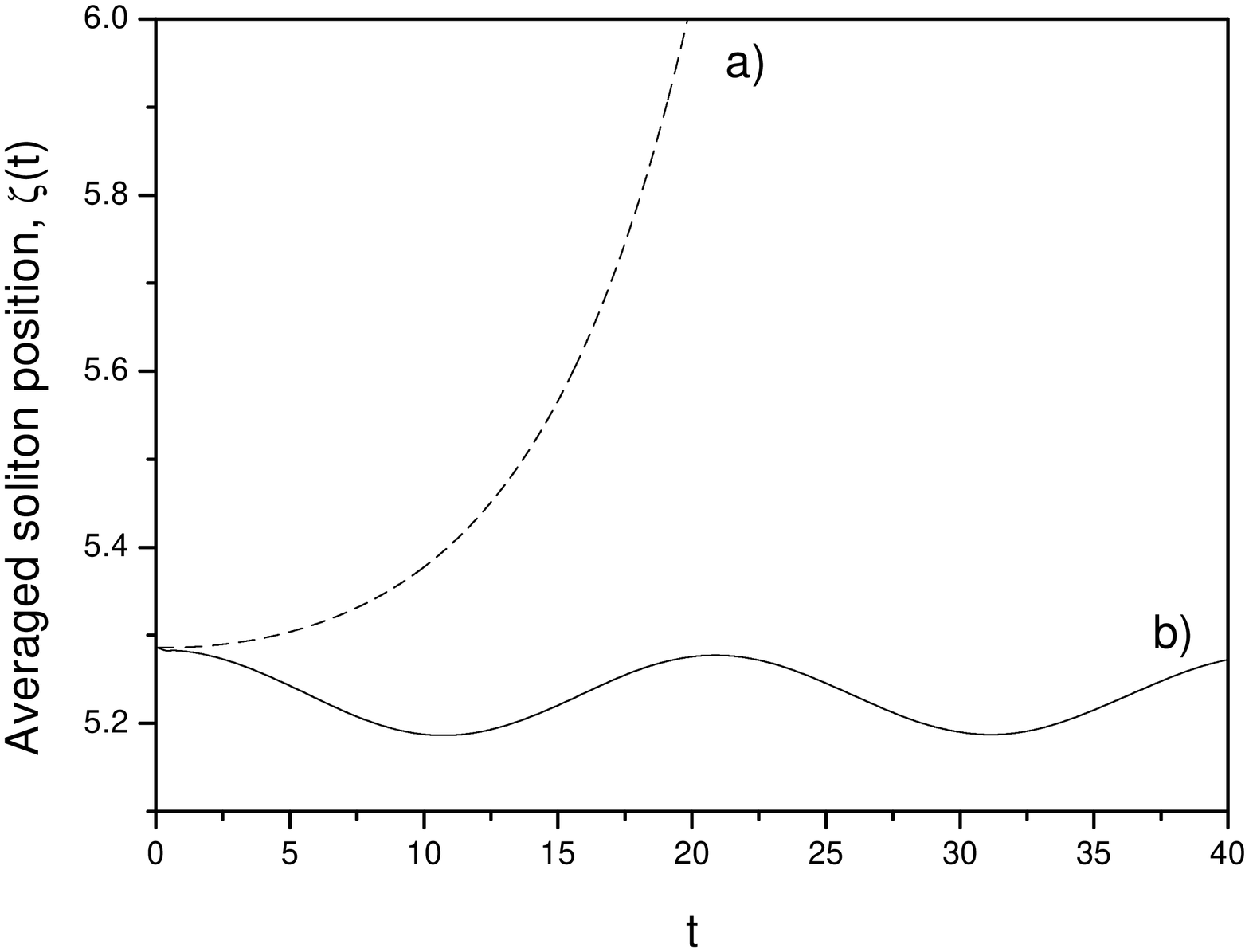,height=3.8in,angle=-90.}}
\end{center}
  \vskip 0.2in
{\protect\small Fig.  \ref{fig_4}. Numerical simulations of the
full GP equation for a spatial periodic potential $f(x) =
\cos(b x)$ with the initial soliton position at an unstable point
$X_{0} = \pi /b$. The case a) (dashed line) corresponds to
infinite motion of the soliton center when $\alpha_{1} = 0$ with
the fast perturbation turned off. The case b) corresponds to
$\alpha_{1} = 9$ when the fast perturbation stabilizes the motion
of the soliton center. The parameters of perturbation are $\omega
= 10$, $\alpha_{0} = 0.05$, as in the case of a bounded potential.
} \vskip0.2in

\noindent

The results of numerical simulation of the full GP equation
are presented in Fig.\ref{fig_4}.
The parameters $\alpha_{0}$ and $\omega$ are the same as for
the above considered bounded potential.
Fig.  \ref{fig_4}a
depicts the dynamics of a soliton center in a periodic potential
$-\alpha_{0} \cos(kx)$ with its initial position
$\zeta = \pi/k$ corresponding to the unstable point
when $\alpha_{1} = 0$ (the fast perturbation is turned off).
As seen the motion is infinite. Fig. \ref{fig_4}b shows the case
when $\alpha_{1} = 9$ (the fast perturbation is turned on).
In this case the motion of the soliton becomes finite and it
oscillates with the frequency $\Omega_{1} = 0.303$ whereas the
predicted value by Eq. (\ref{e16})  is $0.308$. One can see that
the agreement with the theory is remarkable.

\section{Conclusion}

We have investigated the propagation of bright
matter wave soliton  in the Bose-Einstein condensate with
trap potential rapidly
varying in time.

The cases of modulated in time quadratic, bounded  and periodic
trap potentials have been analyzed. For the repulsive  (unstable)
trap potential it is shown that there exists a critical value of
modulation parameters, when the matter wave soliton is
stabilized. Analogous phenomenon of stabilization of unstable fixed
points is found for periodic modulations (optical lattices).
For the bounded potential it is shown that the effective trap
potential can bifurcate from the one well  to the
triple-well structure; and so, may give rise to the splitting of
single attractive BEC
into three parts.

\section{Acknowledgments}
The work is partially supported by Fund of support of the
fundamental research of the Uzbek Academy of Sciences(Award
15-02).  F.A. is grateful to FAPESP for the support of his work.
We are grateful to J.G. Caputo, for the participation at the
initial stage of this work and to R. Kraenkel and L. Tomio
for useful discussions.

\begin{figure}
\caption{Effective potential for $f(x) = 1/\cosh{(bx)}$ after
temporal averaging.
The parameters of perturbation $\omega = 10$, $\alpha_{0} = 0.05$ and
$\alpha_{1} = 9$.
Dashed line is for the case with $\alpha_{1} = 0$ when temporal
perturbation is turned off.}
\label{fig_2}
\end{figure}

\end{document}